# Measure of Synchronism of Multidimensional Chaotic Sequences Based on Their Symbolic Representation in a T–Alphabet

A. V. Makarenko

Constructive Cybernetics Research Group, Moscow, Russia

e-mail: avm.science@mail.ru

**Abstract.** A new approach to analysis of the synchronization of chaotic oscillations in two (or more) coupled oscillators is described that makes it possible to reveal changes in the structure of attractors and detect the appearance of intermittency. The proposed method is based on a symbolic analysis developed previously in the velocity–curvature space of multidimensional sequences and maps. The method is tested by application to a Lorentz system. The results confirm the informativity of the analyzer and reveal specific features of changes in the structure of an attractor of the three-component test system.

**Keywords:** Symbolic Analysis, T–Alphabet, Chaos Synchronization, Multidimensional Systems.

Synchronization of chaotic oscillations [1, 2], which is among the most fundamental concepts of the theory of nonlinear dynamics and chaos, can take place via several routes, including the complete (identical) [3], frequency [4], phase [5], generalized [6], lag [7], timescale [8], and antiphase type [9]. At present, investigations are aimed at (i) considering various types of synchronization from a common standpoint and (ii) seeking for new types of synchronous behavior.

Previously, the author proposed [10] a new method of symbolic analysis that was based on finite discretization of the velocity–curvature space and a minimum alphabet introduced in a natural way. According to this, the sequence $\{\mathbf{s}_k\}_{k=1}^{K}$ ($\mathbf{s} \in \mathrm{S} \subset \mathbb{R}^N$, $k \in \mathrm{K} \subseteq \mathbb{N}$, $K \geq 3$) is encoded by assigning every $n$th component ($n = \overline{1, N}$) the corresponding sequence of terms $\{T_k^{\alpha\varphi}|_n\}_{k=1}^{K}$ [10]. Using the proposed method, it is possible to study in detail the shapes (geometric structures) of trajectories $\{\mathbf{s}_k\}_{k=1}^{K}$ in an $\mathrm{S} \times \mathrm{K}$ space (for this important characteristic, see [11–13] and references therein). The present work has been devoted to constructing a tool for analysis of the synchronism between components in multidimensional sequences with respect to the shape of their trajectories in the $\mathrm{S} \times \mathrm{K}$ space.

Let us consider the $k$th term of sequence $\{\mathbf{s}_k\}_{k=1}^{K}$ completely synchronous in the alphabetic representation $T^{\alpha\varphi}$ provided that:

$$J_{sym}^{\alpha\varphi}\left[T_k^{\alpha\varphi}\right] \equiv 1, \text{ where } J_{sym}^{\alpha\varphi}\left[T_k^{\alpha\varphi}\right] = \begin{cases} 1 & T_k^{\alpha\varphi}|_1 = \ldots = T_k^{\alpha\varphi}|_N; \\ 0 & otherwise. \end{cases} \quad (1)$$

Taking into account the possible antiphase synchronization (antisynchronization) [9], let us also consider the variant of inverting the initial sequence $\{\mathbf{s}_k\}_{k=1}^{K}$ as follows:

$$\{\overline{\mathbf{s}}\}|_m = \left[v_{m,1}\, s_1, \ldots, v_{m,n}\, s_n, \ldots, v_{m,N}\, s_N\right]^{\mathrm{T}}, \; v_{m,1} = +1, \; v_{m,n'} = \pm 1, \; n' = \overline{2, N}, \quad (2)$$

where $\circ^{\mathrm{T}}$ is the symbol of transposition. The correspondence is established as follows [10]:

$$\{\overline{\mathbf{s}}_k\}_{k=1}^{K}|_m \Rightarrow \{T_k^{\alpha\varphi}\}_{k=1}^{K}|_m,$$

where symbols $T^{\alpha\varphi}$ are replaced during inversion of $\overline{\mathbf{s}}_k$ according to the table.

Replacement of $T_k^{\alpha\varphi}|_n$ terms for $s_k^{(n)} \to -1 \cdot s_k^{(n)}$ inversion

| | | | | | | | | | | | |
|---|---|---|---|---|---|---|---|---|---|---|---|
| +1 | T0 | T1 | T2 | T3N | T3P | T4N | T4P | T5N | T5P | T6 | T7 |
| –1 | T0 | T2 | T1 | T5P | T5N | T4P | T4N | T3P | T3N | T7 | T6 |

Taking into account definitions (1) and (2), the integral coefficient of synchronism between components of the sequence $\{\mathbf{s}_k\}_{k=1}^K$ in the alphabetic representation $T^{\alpha\varphi}$ is expressed as follows ($\delta^{\alpha\varphi} \in [0,1]$):

$$\delta^{\alpha\varphi} = \max_m \delta_m^{\alpha\varphi}, \; \delta_m^{\alpha\varphi} = \frac{1}{K}\sum_{k=1}^{K} J_{sym}^{\alpha\varphi}\left[\{T_k^{\alpha\varphi}\}|_m\right]. \quad (3)$$

Thus, according to relations (1)–(3), the proposed analyzer of synchronism in the sequence $\{\mathbf{s}_k\}_{k=1}^K$ estimates the level of complete synchronization [3] in the alphabetic representation $T^{\alpha\varphi}$ – that is, complete synchronization on the level of the terms $\{T_k^{\alpha\varphi}\}_{k=1}^K$ is not imply complete synchronization on the level of sequence $\{\mathbf{s}_k\}_{k=1}^K$. This circumstance opens a potential way to constructively apply the proposed analyzer to investigations of generalized synchronization [6]. In addition, by introducing the operator of shift for $K \to \infty$ as $\mathrm{H}_{\mathbf{h}} : \left\{ T_k^{\alpha\varphi}|_1 \to T_{k+h_1}^{\alpha\varphi}|_1, \ldots, T_k^{\alpha\varphi}|_N \to T_{k+h_N}^{\alpha\varphi}|_1 \right\}$, $h_n \in \mathbb{N} \cup \{0\}$, $h_n << K$, and posing the condition:

$$\delta_m^{\alpha\varphi} = \max_{\mathbf{h}|h_i \to \min} \frac{1}{K+1-k^*}\sum_{k=k^*}^{K} J_{sym}^{\alpha\varphi}\left[\mathrm{H}_{\mathbf{h}}\left(\{T_k^{\alpha\varphi}\}|_m\right)\right], \; k^* = 1 + \min\left(h_1, \ldots, h_N\right), \quad (4)$$

we make the coefficient $\delta^{\alpha\varphi}$ capable of detecting the lag synchronization [7] as well. In this case, the value of $\mathbf{h}^o = \arg_{\mathbf{h}|m} \max_{m|\mathbf{h}} \max_{\mathbf{h}} \delta_m^{\alpha\varphi}$ corresponds to the effective shift in the sequence $\{\mathbf{s}_k\}_{k=1}^K$. These issues will be considered in subsequent investigations.

Coefficient $\delta^{\alpha\varphi}$ defined in (3) characterizes the synchronism between the components of the sequence $\{\mathbf{s}_k\}_{k=1}^K$ on average. However, the adequacy of averaging and the conclusions based on this characteristic have definite limits. For this reason, it is expedient to synthesize characteristics that would allow us to analyze the temporal structure of the synchronism between components of the sequence $\{\mathbf{s}_k\}_{k=1}^K$ in the alphabetic representation $T^{\alpha\varphi}$. With this structure, we indicate manifestations of the intermittency [12, 13], whereby the events of synchronous behavior of the components $\{\mathbf{s}_k\}_{k=1}^K$ are separated by intervals with small values of the coefficient.

Thus, a synchronous domain ($SD$) comprises a set of $T_{k'}^{\alpha\varphi}$ terms from the sequence $\{T_k^{\alpha\varphi}\}_{k=1}^K$, which obeys the following condition:

$$SD_r : \; J_{sym}^{\alpha\varphi}\left[T_{k'}^{\alpha\varphi}\right] \equiv 1, \; J_{sym}^{\alpha\varphi}\left[T_{k''}^{\alpha\varphi}\right] \equiv 0 \vee k'' = 0, \; J_{sym}^{\alpha\varphi}\left[T_{k'''}^{\alpha\varphi}\right] \equiv 0 \vee k''' = K+1, \; \mathrm{SD} \ni SD_r, \quad (5)$$

where $k' = \overline{b_r^{SD}, b_r^{SD} + L_r^{SD}}$, $k'' = b_r^{SD} - 1$, $k''' = b_r^{SD} + L_r^{SD} + 1$. Domain $SD_r$ is characterized by the moment of appearance ($b_r^{SD}$) and spatial length ($L_r^{SD}$) in the $\mathrm{K}$ space, where $r$ is the domain number ($r = \overline{1, R^{SD}}$).

Let us define the domain and subdomain spectral density functions as follows:

$$H^{SD}\left[L^{SD}\right] = \sum_{r=1}^{R^{SD}} \delta[L_r^{SD}, L^{SD}], \quad H^{SS}\left[L^{SD}\right] = \sum_{j=L^{SD}}^{K}\left(j - L^{SD} + 1\right) H^{SD}[j], \quad L^{SD} = \overline{1, K}, \quad (6)$$

where $\delta[\circ,\circ]$ is the Kronecker symbol. These functions bear information about a qualitative structure of the synchronism between components of the sequence $\{\mathbf{s}_k\}_{k=1}^{K}$ in the alphabetic representation $T^{\alpha\varphi}$.

Now let us consider a reference sequence $\xi_{uw} = \{T_k^{\alpha\varphi}\}_{k=1}^{\infty}$ in which the components $\{T_k^{\alpha\varphi}|_{n1}\}_{k=1}^{\infty}$ and $\{T_k^{\alpha\varphi}|_{n2}\}_{k=1}^{\infty}$ are mutually independent, $n1 \neq n2$, $n1, n2 = \overline{1, N}$, and each component has the same probability of realization for every permissible trajectory length $K$.

Main hypothesis $\mathcal{H}_0$. The value of $H^{SD}\left[L^{SD}\right]$ is determined by random coincidence of subsequences in independent components of the analyzed sequence on the confidence level $\alpha$ (permissible error of the first kind [14]), rather than being a consequence of the synchronous behavior of systems. Under this assumption, it is possible to determine some quantities that would allow the synchronism in the analyzed sequences to be revealed to within an error of $\alpha$.

For this purpose, let us introduce the total and filtered maps of synchronization as follows:

$$M_k^{SD} = \begin{cases} L_r^{SD} & k \in \left[b_r^{SD}, b_r^{SD} + L_r^{SD}\right]; \\ 0 & otherwise. \end{cases}, \quad M_{\alpha k}^{SD} = \begin{cases} M_k^{SD} & M_k^{SD} > L_\alpha^{SD}; \\ 0 & otherwise. \end{cases}, \tag{7}$$

where $L_\alpha^{SD}$ is the boundary size of the synchronous domain, for which the probability of at least a single appearance of longer domains in the reference sequence $\xi_{uw}$ of length $K$ does not exceeds the confidence level $\alpha$. Note that $M^{SD}$ and $M_\alpha^{SD}$ maps provide a transition from the structure of synchronism in synchronous domain directly to that in space K. In the case of sequences with pronounced statistical proper ties defined over $\mathbb{R}^N \times \mathbb{N}$ and/or with a degenerate set of $T^{\alpha\varphi}$ terms for $Q^{\alpha\varphi}$ transitions (see, e.g., [10]), it is necessary to change the reference sequence that is used to calculate the parameters of various statistical hypotheses related to the analysis of synchronism [14].

The proposed approach to determination of the level of synchronism between components of a sequence has been used to study oscillations in the model Lorentz system [15–18]:

$$\dot{x} = -\sigma(x-y),\ \dot{y} = r\,x - y - x\,z,\ \dot{z} = -\beta\,z + x\,y, \quad \mathbf{s}(t) = [x(t), y(t), z(t)]^{\mathrm{T}}. \tag{8}$$

Note that analysis of characteristics of the symbolic synchronization between components of trajectory $\mathbf{s}(t)$ can be treated in two ways, namely, as investigation of (i) the shape of an attractor of system (8) or (ii) the synchronization of coupled structurally nonidentical systems [2] (Fig. 1a).

In numerical simulations, the values of $\sigma = 10$ and $\beta = 8/3$ were fixed and parameter $r$ was varied on the interval $r \in [20, 300]$ at a step of $\Delta r = 0.5$. Equations (8) were integrated by the RADAU5 method on the interval of $T = [0, 300]$ at $\Delta t = 2.5\times10^{-3}$. For every $r$ value, 60 trajectories have been calculated for a set of initial conditions $x_0 = \xi_1 \in [-10,10]$, $y_0 = \xi_2 \in [-10,10]$, $z_0 = \xi_3 \in [0,10]$, where $\xi_{1-3}$ are noncorrelated uniformly distributed random values. This approach reduced to a minimum the memory effect induced by the initial condition. For each trajectory $\mathbf{s}(t)$, the sequence $\{T_k^{\alpha\varphi}\}_{k=1}^{K}$ of length $K = 2\times10^4$ (interval of $T' = [250, 300]$) has been generated by means of stroboscopic Poincare mapping [15]. The shift from $t = 0$ is explained by the need for neutralization of a parasitic effect related to the transient process. Note that the interval of parameter $r$ under consideration includes attractors of two types [15, 16]: quasi-hyperbolic ($r_a = 28$) and nonhyperbolic ($r_b = 210$).

Figures 1b–1d show that the dependence of coefficient $\delta^{\alpha\varphi}$ on $r$ is significantly influenced by the selected set of $\mathbf{s}(t)$ components. In particular, the value of $\delta^{\alpha\varphi}$ for the ($x, y$) pair reaches a maximum at $r = 23.5$ and then exhibits a monotonic decrease. For the ($x, z$) pair, the coefficient $\delta^{\alpha\varphi}$ increases with $r$. For both these pairs (as well as for the $x, y, z$ triad) the values of $\delta^{\alpha\varphi}$ within the confidence interval are significantly greater than those for the reference sequence $\xi_{uw}^{Te}$. This fact is indicative of a nonaccidental synchronization between $\mathbf{s}(t)$ components in system (8) under consideration. A minimum value of $\delta^{\alpha\varphi}$ is observed for the $(y,z)$ pair at $r < r_{ew2}$, $\mathrm{M}\left[\delta^{\alpha\varphi}(y,z)\right] < M_{\xi 2}$. It should also be noted that there are two windows, $[r_{bw1}, r_{ew1}]$ and $[r_{bw2}, r_{ew2}]$, in which the value of $\delta^{\alpha\varphi}$ for the $(x,y)$, $(x,z)$, and $x,y,z$ sets is stabilized (see Figs. 1b and 1d) and its variance significantly decreases, which is evidence for a robustness of the synchronization with respect to a change in the initial conditions. On the other hand, there are points and intervals (e.g., $r_{1-3}$, $r_{e1}$, $[r_{pb}, r_b]$) where $\mathrm{D}\left[\delta^{\alpha\varphi}\right]$ significantly increases as compared to that in some other intervals $r \in [r_a, r_b]$.

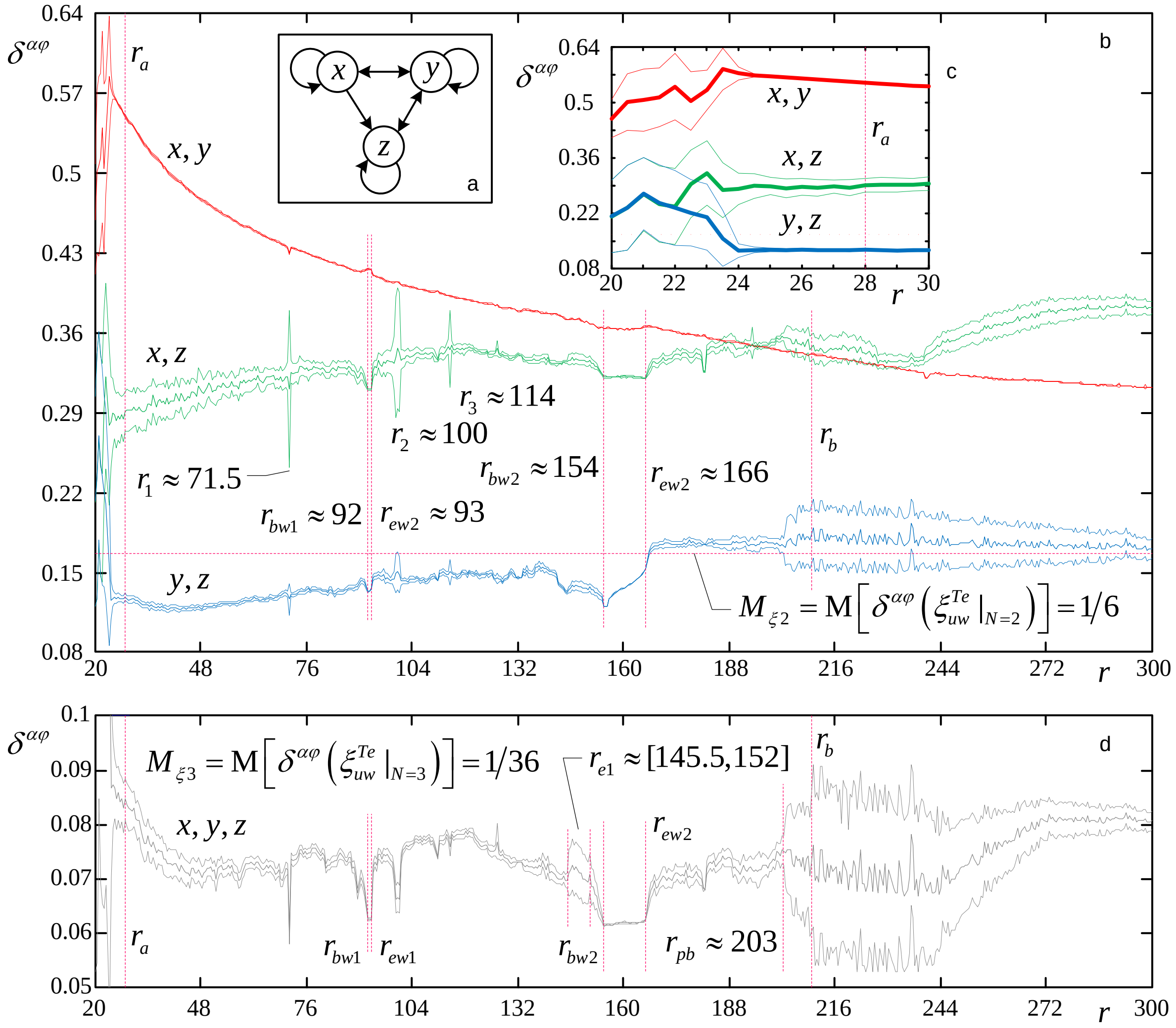

**Fig. 1.** Analysis of oscillations in model system (8): (a) schematic diagram of coupling between variables; (b, c) plots of coefficient $\delta^{\alpha\varphi}$ vs. $r$ for $\mathbf{s}(t)$ sets with two components; (d) the same for $\mathbf{s}(t)$ sets with three components. Confidence intervals are indicated for a probability of $1-\alpha$ ($\alpha = 10^{-3}$); the reference sequence as formed in the alphabet $T_e^{\alpha\varphi} = \{\mathrm{T3N}, \mathrm{T3P}, \mathrm{T5N}, \mathrm{T5P}, \mathrm{T6}, \mathrm{T7}\}$ (see [10]).

Variation of parameter $r$ in the general case leads to a change in the form of a synchronizing configuration of $\mathbf{s}(t)$ components selected on the manifold of initial conditions, which conforms the importance of an analysis for the antisynchronization [9]. The sets of phase variables $(x,z)$ and $(y,z)$ are represented by a rather regular switching of configurations $xz \leftrightarrow x\bar{z}$ and $yz \leftrightarrow y\bar{z}$, while the $(x,y)$ set is invariant with respect to the variation of $r$ and is always represented by the $xy$ configuration. Figure 2 shows the form of a synchronizing configuration for the set of three components, which reveals a special window of $[r_b, r_{bp}]$.

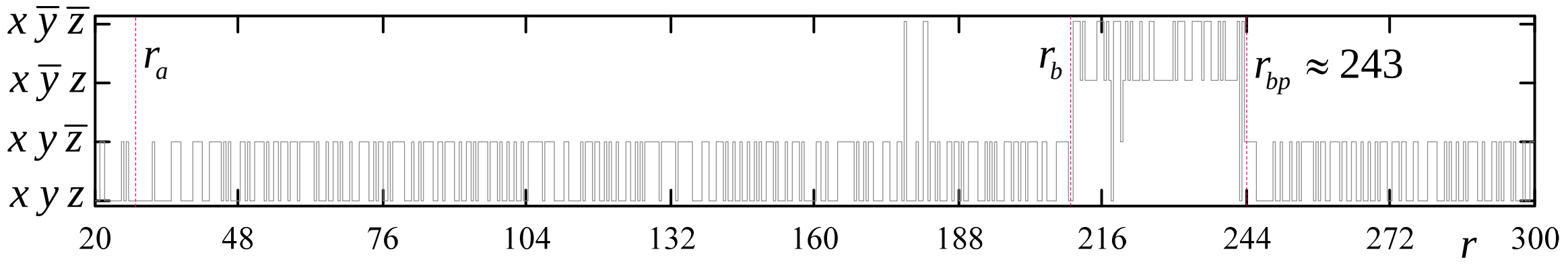

**Fig. 2.** The form of a synchronizing configuration for the $(x, y, z)$ set as dependent on parameter $r$; component (2) with $v_{m,n'} = -1$ is indicated by the bar $\bar{\circ}$.

Figure 3 shows that the domain structure of synchronization in map (8) for the $(x, y, z)$ set mapped as $L^{SD}$ is nontrivial and contains outbursts (e.g., at $r_4$), broadenings (e.g., $[r_b, r_{bp}]$), and forbidden vales (e.g., $[r_{bw1}, r_{ew1}]$). Analysis also shows significant manifestations of intermittency [12, 13] in the synchronization of $\mathbf{s}(t)$ components.

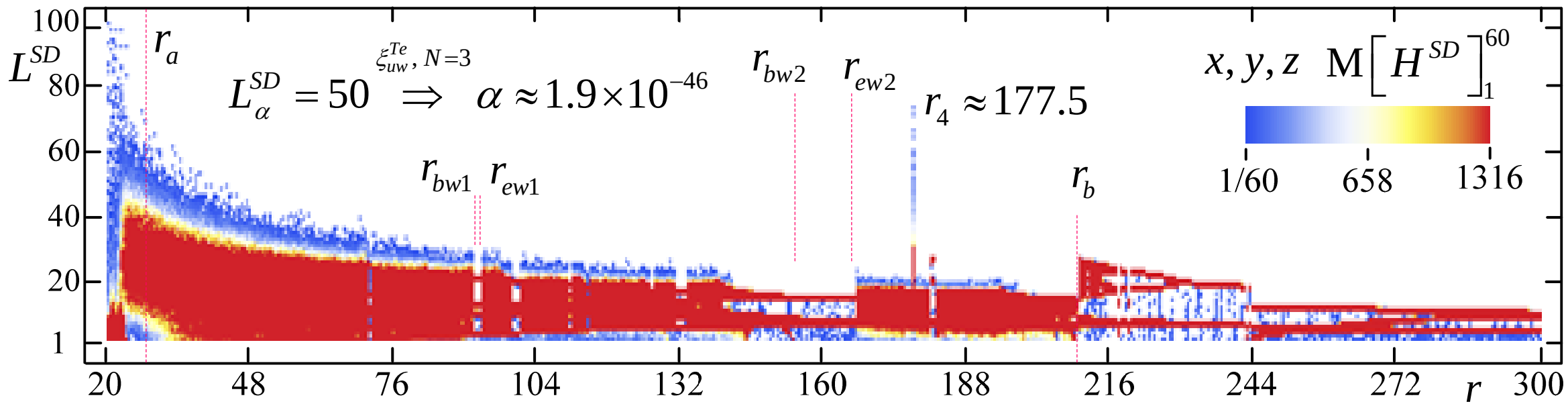

**Fig. 3.** Map of $L^{SD} : H^{SD}\left[L^{SD}\right] > 0$ as dependent on parameter $r$; $L^{SD}_\alpha = 6$ for $\xi^{Te}_{uw}$ at $\alpha = 10^{-3}$ and $K = 2\times10^4$.

Thus, a new approach to quantitative evaluation of the level and parameters of the synchronization of chaotic oscillations in two or more coupled oscillators has been described that makes it possible to reveal changes in the structure of attractors and detect the appearance of intermittency. The synchronization is estimated from the standpoint of the shape (geometric structures) of trajectories in the $\mathrm{S}\times\mathrm{K}$ space, that is, topological synchronism of dynamical systems. The proposed method has been tested by application to a Lorentz system. The results confirm the informativity of the analyzer and reveal specific features of changes in the structure of an attractor of the three-component test system. It is planned to check the obtained results for other values of $\Delta t$ and relate them to the structural and dimensional features of attractors of system (8). Subsequent studies are intended to expand the analytical possibilities of the proposed approach.

*Translated by P. Pozdeev*

**Andrey V. Makarenko** – was born in 1977, since 2002 – Ph. D. of Cybernetics. Founder and leader Research & Development group "Constructive Cybernetics". Author and coauthor of more than 50 scientific articles and reports. Associate Member IEEE (IEEE Systems, Man, and Cybernetics Society Membership). Research interests: analysis of the structure dynamic processes, predictability; detection, classification and diagnosis is not fully observed objects (patterns); synchronization in nonlinear and chaotic systems; system analysis and modeling of economic, financial, social and bio-physical systems and processes; system approach to development, testing and diagnostics of complex information-management systems.